\documentclass[conference, 10pt]{IEEEtran}
\usepackage{amsfonts}
\usepackage{amsmath}
\usepackage{amsthm}
\usepackage{amssymb}
\usepackage{graphicx}
\usepackage{subfigure}
\usepackage{algorithm}
\usepackage{algorithmic}

\begin{document}
%
\title{Hybrid Radio-map for Noise Tolerant Wireless Indoor Localization}


%
\author{\IEEEauthorblockN{Xiongfei Geng\IEEEauthorrefmark{1},
Yongcai Wang\IEEEauthorrefmark{2},
Haoran Feng\IEEEauthorrefmark{3} and
Zhoufeng Chen\IEEEauthorrefmark{1} }
\IEEEauthorblockA{\IEEEauthorrefmark{1}China Waterborne Transport Research Institute, Beijing, P. R. China}
\IEEEauthorblockA{\IEEEauthorrefmark{2}Institute for Interdisciplinary Information Sciences (IIIS),  Tsinghua University, Beijing, P. R. China}
\IEEEauthorblockA{\IEEEauthorrefmark{3}National Engineering Research Center of Software Engineering, Peking University, Beijing, P. R. China \\ 
}}


\maketitle

\begin{abstract}
In wireless networks, radio-map based locating techniques are commonly used to cope the complex fading feature of radio signal, in which a radio-map is built by calibrating received signal strength (RSS) signatures at training locations in the offline phase. However, in severe hostile environments, such as in ship cabins where severe shadowing, blocking and multi-path fading effects are posed by ubiquitous metallic architecture, even radio-map cannot capture the dynamics of RSS. In this paper, we introduced multiple feature radio-map location method for severely noisy environments. We proposed to add low variance signature into radio map. Since the low variance signatures are generally expensive to obtain, we focus on the scenario when the low variance signatures are sparse. We studied efficient construction of multi-feature radio-map in offline phase, and proposed feasible region narrowing down and particle based algorithm for online tracking. Simulation results show the remarkably performance improvement in terms of positioning accuracy and robustness against RSS noises than the traditional radio-map method. 
\end{abstract}

\IEEEpeerreviewmaketitle

\section{Introduction}

For on-ship wireless sensor networks \cite{paik_characteristics_2009,  kdouh_double_2013,mu_hybrid_2012}, one of the most important and desirable applications is to provide real-time location information for crews, passengers or facilities by using the sensor network as a wireless locating infrastructure. When the ships seal on the sea, the location information will be fundamental context for safety-oriented applications or on-ship facility management.  

For locating using sensor networks, wireless location techniques have attracted great research attentions in recent years. Various locating techniques have been developed using  different implementation techniques, but it is still challenging to find a balance between the positioning accuracy and the system cost. Some positioning techniques provide good positioning accuracy, such as TOA (Time of Arrival) \cite{zhao_autonomous_2008} or TDOA (Time Difference of Arrival)\cite{gustafsson_positioning_2003} based localization methods, but these methods generally need special hardware, such as ultrasound or acoustic transducers, which need additional hardware costs. Some other methods are inexpensive, such as RSS (Radio Signal Strength) based wireless location \cite{el-kafrawy_propagation_2010}, because RSS information is free-of-charge. But these inexpensive methods provide only coarse-grained  positioning accuracy.  

A way to increase the positioning accuracy by using RSS is to exploit a \emph{radio-map} based locating method \cite{haque_profiling-based_2013, scholl_fast_2012}, which trains the RSS \emph{fingerprints} of all cared locations in an offline calibration phase,  to construct a \emph{radio map};  then the online measured RSS of the target is searched in the radio-map to find the location whose RSS signature matches best to the online RSS measurement. The location is chosen as the position estimation of the target. Probabilistic radio-maps and Bayesian reasoning methods can be applied to improve the positioning accuracy. 

In our previous works, we have studied both ultrasound TOA-based locating systems \cite{zhao_autonomous_2008, wang_lock:_2009} and radio-map based locating systems in buildings \cite{zheng_hips:_2009}. But when we developed and tested these systems on ships, dramatic performance degradation was found because of the signal blocking effects by the metallic architecture of the cabins. The multi-path,  shadowing, and blocking effects are serious.  The received signal strength can be very weak even when the receivers are close to a transmitter but are not in line of sight (NLOS).  To deal with this problem, in this paper, we propose efficient methods to construct multi-feature radio-map to overcome the hostile environments for wireless localization.  

To be tolerant to the noise of RSS, we proposed hybrid radio-map integrating both low-variance signature and the RSS signature. More particularly, we deploy ultrasound beacons sparsely in the sensing field which contributes sparse, by low-variance time-of-arrival (TOA) information of ultrasound from the transmitter to receiver. We show that by integrating these sparse low variance information into radio-map, it can dramatically improve the positioning accuracy of Radio-map based positioning systems.  

Utilizing the lower variance signature, we proposed a new efficient method to offline calibrate  hybrid radio-maps without the pain of manual calibration.  Then in the online phase, instead of simple matching algorithm, we propose to use the low variance signature to narrow down the feasible space firstly, and then use particle filter based algorithm to efficiently and accurately tracking the mobile targets. Since the low variance beacons are very sparse, only a little additional costs are needed, but the hybrid radio-map system can provide dramatical improvement in positioning accuracy and reliability. 

The remainder sections are organized as following. Background and related works are introduced in Section 2. We introduce the hybrid radio-map model construction and online tracking algorithm in Section 3. Simulation based evaluation results are introduced in Section 4. Conclusions are drawn in Section 5.  

\section{Background and Related Works}
Using wireless networks as indoor locating infrastructure, there are different ways to utilize the wireless signal. One way is to use the propagation model of RF signal as a ranging reference. Some research works studied the RF attenuation model in indoor environments \cite{el-kafrawy_propagation_2010}, so that distances from a target to a set of beacons can be inferred from the amount of RF attenuations. Then least square estimation or multilateration methods \cite{han_localization_2013} are applied to the distance set to estimate the position of the target. Although this method is simple to calculate, the positioning accuracy is coarse,  because even empirical propagation model cannot capture the dynamics of indoor environments.     
\subsection{Radio-map Locating Method}
To improve the positioning accuracy, \emph{pattern-matching} based approach was proposed to model the diverse fading signatures of radio signal \cite{haque_profiling-based_2013}\cite{ni_landmarc:_2003}\cite{bahl_radar:_2000}. This method contains an offline and an online phase. In offline phase, $n$ training locations are selected in the sensing field, which are denoted by $\mathbf{L}=\{l_{1}, l_{2}, \cdots, l_{n}\}$.  Suppose there are $m$ beacons (WiFi APs or wireless sensors) in the sensing field, which are denoted by $\mathbf{B}=\{b_{1}, b_{2}, \cdots, b_{m}\}$.  In the training phase, the RSS values of all beacons at each training location $l_{i}$ will be measured over a period of time, so that a \emph{signal signature} vector of location $l_{i}$ is constructed as $\mathbf{r}_{i}=\{r_{i,1}, r_{i,2}, \cdots, r_{i,m}\}$. When only mean value of RSS is considered, $r_{i,j}$ represents the average RSS value from $b_{j}, j=1, \cdots, m$. When signature distribution is considered,  $r_{i,j}$ can be probabilistic density function (pdf) of RSS from $b_{j}$.  The signature vectors of all training locations are stored as a database,  called \emph{radio-map}, denoted by $\mathbf{R}=\{\mathbf{r}_{1}, \mathbf{r}_{2}, \cdots, \mathbf{r}_{n}\}$.  

In the online positioning phase, a mobile target measures its current RSS vector $\mathbf{s}=\{s_{1}, s_{2}, \cdots, s_{m}\}$ and finds the best match (Euclidean distance in signal space) of $\mathbf{s}$ in $\mathbf{R}$ to estimate the position of the target. In mean value type radio-map, matching can be conducted by Nearest Neighbor algorithm\cite{bahl_radar:_2000}. In pdf type radio-maps, maximum likelihood estimation and Bayesian estimation can be applied. 
When radio-map is trained in fine granularity and the environments are not highly dynamic, the positioning accuracy of radio-map based method can be in 2-3 meters resolution. 

But the positioning accuracy may become worse in hostile environment such as in ship cabins, where the shadowing and multi-path fading effects are severe and the RSS signatures change over time.  Another problem is that the radio-map calibration process is general time consuming and laborious, which generally needs deliberate training method \cite{scholl_fast_2012}. 
\subsection{Locating by Time of Arrival (TOA)} 
A more accurate approach is to utilize the speed difference of signal propagation to measure distances from transmitters to receivers, so as to conduct indoor locating more accurately\cite{zhao_autonomous_2008}\cite{gustafsson_positioning_2003}. Ultrasound and acoustic signals are the generally exploited low speed signals. In the case of measuring time of arrival, the transmitter broadcasts low-speed signal (ultrasound or acoustic) and RF signal simultaneously. The receiver receives the RF signal to synchronize timer with the transmitter and then measures the traveling time of the low-speed signal to estimate distance from the transmitter. Only when a receiver $j$ is within the communication range of the low speed signal (denoted by $R$, and generally small) of the transmitter $i$, can a distance $d_{i,j}$ be measured. When a set of distances, which is denoted by $\mathbf{D}_{i}=\{d_{i,j}\}$ are obtained, least square estimation or multilateration is applied for position calculation. TOA-based positioning can provide centimeter level positioning accuracy\cite{wang_lock:_2009}. But because the short transmission range of the low speed signals (ultrasound and acoustic), and the requirement of more than three non-collinear distances for location estimation, TOA-based positioning requires dense deployment of TOA beacons, which poses high cost to the positioning system.

\section{Hybrid Radio-Map Locating Method}
Note that the radio-map based and TOA-based wireless locating methods both have advantages and shortcomings. We propose a method to integrate their advantages and to avoid their shortcomings.  Our proposed method is not specifically designed for integrating RSS and TOA signatures, it is actually designed for integrating RSS with a low variance signature such as TOA, time difference of arrival (TDOA) etc.  Therefore, in the following model, we call the second signature low variance signature (LVS). 

Let's consider a hybrid positioning system containing a set of RF beacons and some sparsely deployed LVS beacons in the sensing field. The RF beacons are denoted by $\mathbf{B}=\{b_{1}, b_{2}, \cdots, b_{m}\}$ and the LVS beacons are denoted by $\mathbf{V}=\{v_{1}, v_{2}, \cdots, v_{g}\}$. 

In offline phase, some training locations $\mathbf{L}=\{l_{1}, l_{2}, \cdots, l_{n}\}$ are selected in the sensing field. At each training location, after a training target listens to beacon signals for a period of time. It can learn a set of beacon signatures $\mathbf{s}_{l}=\{d_{1,l}, \cdots, d_{g,l}, s_{1,l}, \cdots, s_{m,l}\}$, where $d_{i,j}$ and $s_{i,j}$ are the LVS signature from $v_{j}$ and RSS signature from $b_{j}$ respectively.  We consider the signatures are calculated by taking average on the collected signatures in the training time. We assume the LVS signature is distance-based signature, such as TOA or TDOA. An identical variance $\delta$ is assumed for all the LVS signatures. We store the multi-feature vector as the signature of location $l$. 

In online phase, a target can online detect a set of beacon signals $\mathbf{s}^{'}=\{d_{1}^{'}, \cdots, d_{g}^{'}, s_{l}^{'}, \cdots, s_{m}^{'}\}$.  Note that, for the limited communication range of beacons, many entries of $\mathbf{s}^{'}$ are zero. We design efficient algorithm to match $\mathbf{s}^{'}$ against the radio map to find a location $l'$ whose radio signature has the least distance to  $\mathbf{s}^{'}$ as the position estimation of the target. 

Since the mobile target has limited moving speed, its historical track implies important clues for its future position. Therefore, based on the online hybrid radio-map locating scheme, we designed particle filter algorithm to more accurately track the movements of the mobile targets.

\begin{figure}[t]
\begin{center}
\includegraphics[width=3.0in]{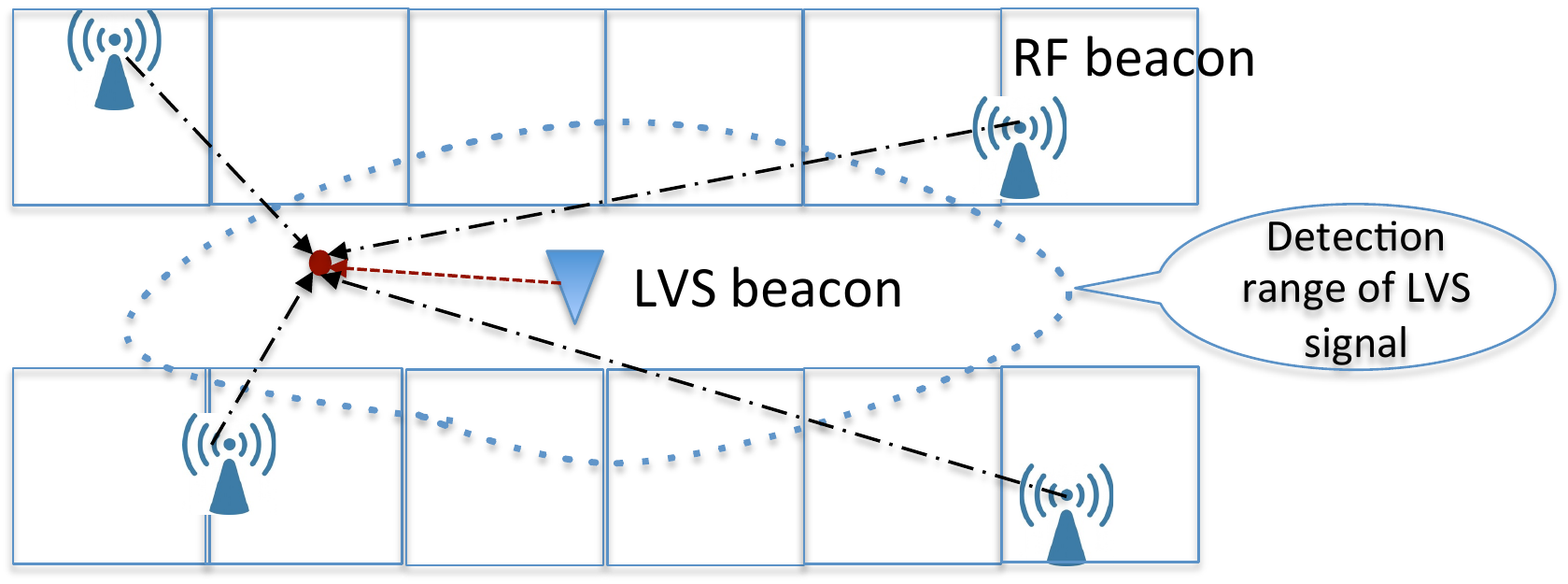}
\caption{An example of hybrid indoor locating system}
\label{fig1}
\end{center}
\end{figure}

We use example in Fig.\ref{fig1} to illustrate the scenario of  hybrid indoor locating system.  The dashed curve is the communication range of the LVS beacons. Note that the communication ranges of beacons are irregular in indoor environments. 

\begin{figure}[t]
\begin{center}
\includegraphics[width=3.2in]{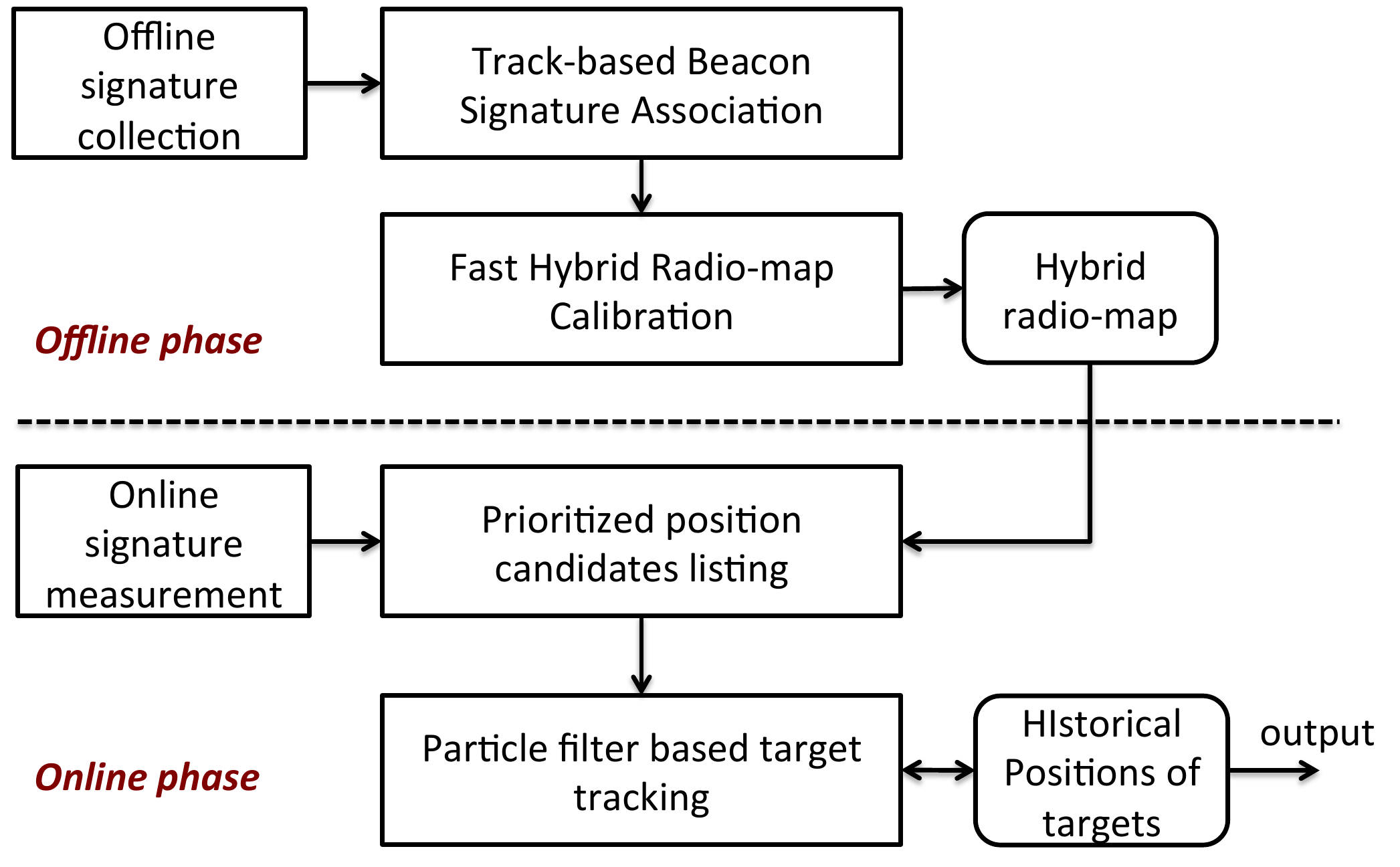}
\caption{Routine of proposed methods for hybrid indoor locating system}
\label{fig2}
\end{center}
\end{figure}

The overview of this paper for calibrating, locating and tracking using hybrid radio-map is shown in Fig.\ref{fig2}.  With the advantage of hybrid beacons, we at first present a method to fast calibrate the hybrid radio-map. It is based on the controlled tracks of the target movements. Then in the online phase, we presented an algorithm to assign radio signature different priorities to list a set of possible positions of the targets by matching the online measured signatures of targets against the radio map. We then present particle filter based target tracking algorithm to utilize the historical position information to accurately track the mobile targets.  

\subsection{Fast Calibration of Hybrid Radio-map}
Radio-map calibration is known time-consuming and laborious, which is a major limitation of radio-map based locating.  By utilizing the advantages of hybrid beacons, we present an efficient method for fast radio-map calibration. 

The LVS beacon, whether using TOA or TDOA signature can provide ranging information from the target to the beacon with good precision. So that in the offline calibration phase, instead of manually assigning location labels to the signal signatures, we design controlled tracks for the training targets to collect hybrid signatures and then use algorithm to calculate the location labels for the hybrid signatures.  The working flow of fast calibration method is as following:
 \begin{enumerate}
\item \emph{Feature points selection:} we select some feature points in the sensing field, which are denoted $\mathbf{F}=\{f_{1}, f_{2}, \cdots, f_{h}\}$. Let $h$ denote the number of feature points. The rule to select the feature points is that the path between two neighboring feature points is a directed line. We manually assign location labels to the hybrid signatures of these feature points.    
\item \emph{Controlled training paths:} A training target is moved along the paths that connecting the feature points. Each segment of its movement is a line, and the whole path can be represented by a set of feature points $\mathbf{p}_{i}=\{f_{i,1},f_{i,2},\cdots, f_{i,e_{i}}\}$, where $f_{i,j}\in \mathbf{F}, j=1, \cdots, e_{i}$. Since we know the start point and end point of each line segment, we can infer the location labels of all the intermediate points.  
\item \emph{LVS-aided fast calibration: } to calibrate the location label of each point on the line segment, we adopt two methods based on whether the point is in the communication range of the LVS beacon.  
\end{enumerate}

\textbf{Case 1}: When the point whose location label need to be calibrated is in the communication range of a LVS beacon, since we know the starting point and ending point of this line segment and know the distance from this point to a LVS beacon (provided by the LVS signature), we can calculate the location label of this point precisely.  Let $(x_{s}, y_{s})$ and $(x_{e}, y_{e})$ be the coordinates of the starting point and end point of the line segment that the point is on. The problem of calibrating the location label of the point is to find a point on this line who has distance $d_{i}$ to the LVS beacon $v_{i}$. This point can be calculated by:
\begin{equation}
\left\{ {\begin{array}{*{20}{c}}
{y = \frac{{{y_e} - {y_s}}}{{{x_e} - {x_s}}}x + \frac{{{y_s}{x_e} - {y_e}{x_s}}}{{{x_e} - {x_s}}}}\\
{\sqrt {{{(y - {y_{{v_i}}})}^2} + {{(x - {x_{{v_i}}})}^2}}  = {d_i}}
\end{array}} \right.
\label{eq1}
\end{equation}

\begin{figure}[htbp]
\begin{center}
\includegraphics[width=2.5in]{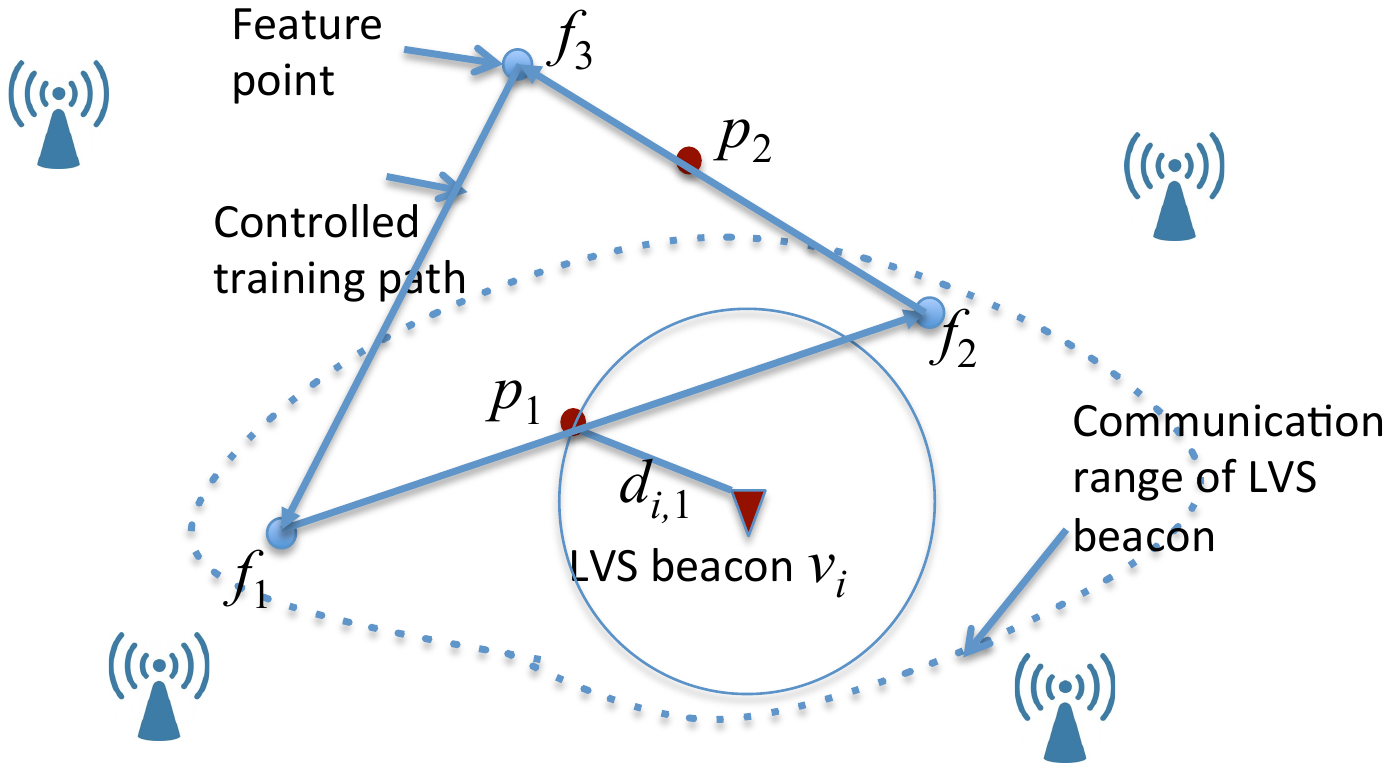}
\caption{Example of LVS-aided fast calibration method.}
\label{fig3}
\end{center}
\end{figure}
This equation array generally contains two ambiguous solutions. But from the moving direction of the target, we can easily disambiguate to determine an unique solution. As shown by the example in Fig.\ref{fig3}, $p_{1}$ is a point on a line segment that is in the communication range of a LVS beacon. We want to assign location label to the RSS signatures measured at this point. There are two intersecting points of  the line with the circle centered at $v_{i}$ with radius $d_{1}$. They are both solutions of Eqn.(\ref{eq1}). But since the target is moving from $f_{1}$ to $f_{2}$, the first point with distance equal to $d_{1}$ is the location of $p_{1}$, which disambiguate the problem.

\textbf{Case 2}: When the point whose location label need to be calibrated is out the communication range all LVS beacons, because we know the starting point and ending point of the line segment, we can infer its location reasonably by linear interpolation. Suppose the training target moves in constant speed. If the times at the starting point, ending point and the point to be calibrated are $t_{s}, t_{e}, t_{p}$ respectively, then the location of the point can be estimated by $\left( {{x_s} + \frac{{({t_p} - {t_s})({x_e} - {x_s})}}{{{t_e} - {t_s}}},{y_s} + \frac{{({t_p} - {t_s})({x_e} - {x_s})}}{{{t_e} - {t_s}}}} \right)$. $p_{2}$ in Fig.\ref{fig3} is such a case. Its position can be inferred by interpolation based on the location of $f_{2}$ and $f_{3}$. 

\subsection{Prioritized Online Positioning Algorithm}
After the hybrid radio-map is calibrated in the offline phase, the system turns to online phase to track the positions of mobile targets. A target can detect a set of beacon signals $\mathbf{s}^{'}=\{d_{1}^{'}, \cdots, d_{g}^{'}, s_{1}^{'}, \cdots, s_{m}^{'}\}$. Because of the sparse deployment of the LVS beacons and there limited communication range, most of the LVS entries are zero. In case the target is out the communication range of all the LVS beacons, all the LVS entries are zero. 

Because the LVS signature has much less variance than that of the RSS signature, only if there is one LVS signature in $\mathbf{s}^{'}$, the LVS signature will provide very valuable information for target position estimation. Instead of simply matching the online measured signature in the radio-map, we present a prioritized approach to always process the LVS signatures first. 
\begin{enumerate}
\item \emph{Characterize feasible region by LVS signature.} A non-zero LVS signature $d_{i}^{'}$ indicates that the distance from the target to $v_{i}$ is a random variable with distribution $N(d_{i}^{'}, \delta)$. Since $\delta$ is small, we can think the target is on a circle with distance at most $d_{i}^{'}+3\delta$, at least $d_{i}^{'}-3\delta$ around the beacon $v_{i}$. This region is called \emph{feasible region}, where the target must locate in. The feasible region can dramatically narrow down the searching space for target position. 
\item \emph{Find possible locations of target in the feasible region.} In the second step, the online measured RSS signature is compared to the trained RSS signature of locations in the feasible region. The locations whose trained RSS signatures match well with the online measurement is elected as possible positions of the  target. Let $\mathbf{L}^{e}$ denote the possible positions of the target. Let set $\mathbb{F}$ include the locations in the feasible region. For all $ l \in \mathbb{F}$, $l$'s trained RSS signature is compared to the RSS signature in $\mathbf{s}^{'}$. If $\left\| {{s_{i,l}} - {{s'}_i}} \right\|_2 <H$, i.e., the RSS distance is less than a threshold, $l$ is added into the possible position set $\mathbf{L}^{e}$. 
\begin{equation}
{L^e} =  \cup l,{\rm{ }}\forall l \in \mathbb{F}{\rm{, if }}\left( {\sum\limits_{i = 1}^m {{{\left\| {{s_{i,l}} - {{s'}_i}} \right\|}_2}}  < H} \right)
 \end{equation}
$H$ is a RSS-distance threshold which rules out the locations whose RSS signatures don't match RSS signature of $\mathbf{s}^{'}$.
\end{enumerate}

\begin{figure}[htbp]
\begin{center}
\includegraphics[width=3.0in]{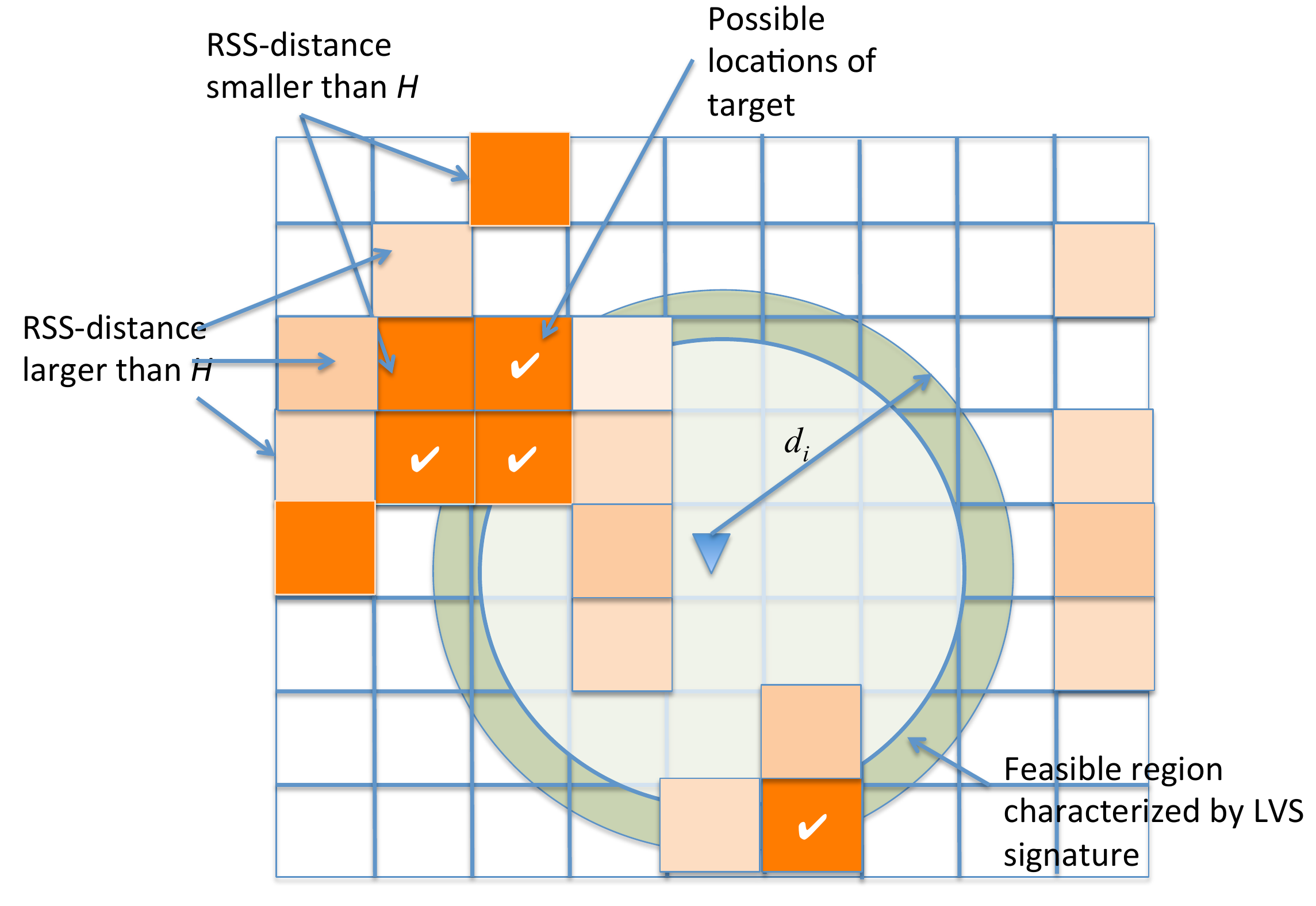}
\caption{Example of prioritized online positioning algorithm.}
\label{fig4}
\end{center}
\end{figure}

An example of the prioritized positioning process is shown in Fig.\ref{fig4}. The green circle is the feasible region characterized by the LVS signature. The color in the map shows whether the RSS-distance of the location is smaller than $H$. Only the locations in brown color have RSS-distance smaller than $H$. Among them, only the locations which are also in the feasible region are added into the possible target position set, which are marked by a ``\checkmark''.

\subsection{Particle Filter based Target Tracking}
Note that the online positioning algorithm provide a set of possible positions instead of a unique position estimation. It is designed in this way because the RSS signature is unreliable. If we determine a unique location using the RSS signature, large locating error maybe incurred. Therefore, we keep a set of possible positions at each time and proposed particle filtering based algorithm to find the optimal moving track of the target.  Let's denote the possible locations of the target calculated at time $t$ is in location set $\mathbb{L}(t)$. 
\subsubsection{Particle Filter}
Since every target has an ID, we only need to consider the case of tracking one target. At any time $t$, we generate $K$ particles (or candidate trajectories), in the possible locations of the target. Let's denote the $k$th particle $z_{k}[t]$. At the next time instant $t+1$, we generate $m>K$ position candidates uniformly at random in $\mathbb{L}(t)$ for $z_k[t+1]$. We now have $mK$ candidate trajectories (particles). Pick the $K$ particles with the best cost functions to get the set ${z_{k}[t+1], k = 1, …,K}$, where the cost function is to specified shortly. Repeat until the end of the time interval of interest. The final output
is simply the particle (trajectory) with the best cost function.

\subsubsection{Cost Function} The cost function is designed based on the fact that the mobile target has restriction in its moving speed (a target will not change speed suddenly). Therefore, we proposed an cost function that penalizes changes in the vector velocity. When a candidate position $z_{k}(t+1)$ is chosen from the current possible position set $\mathbb{L}(t+1)$,  the increment in position $z_{k}(t+1)-z_{k}(t)$ is an instantaneous estimate of the velocity vector at time $t$. The cost at time $t$ is therefore defined as the norm squared of the difference between the velocity vector estimates at time $t$ and $t-1$. This is:
\begin{equation}
\begin{array}{c}
{c_k}[t] = {\left\| {\left( {{z_k}[t + 1] - {z_k}[t]} \right) - \left( {{z_k}[t] - {z_k}[t - 1]} \right)} \right\|^2}\\
 = {\left\| {{z_k}[t + 1] + {z_k}[t - 1] - 2{z_k}[t]} \right\|^2}
\end{array}
\end{equation}
\begin{figure*}[t]
  \begin{minipage}[t]{0.33\linewidth} 
    \centering 
    \includegraphics[width=1.7in]{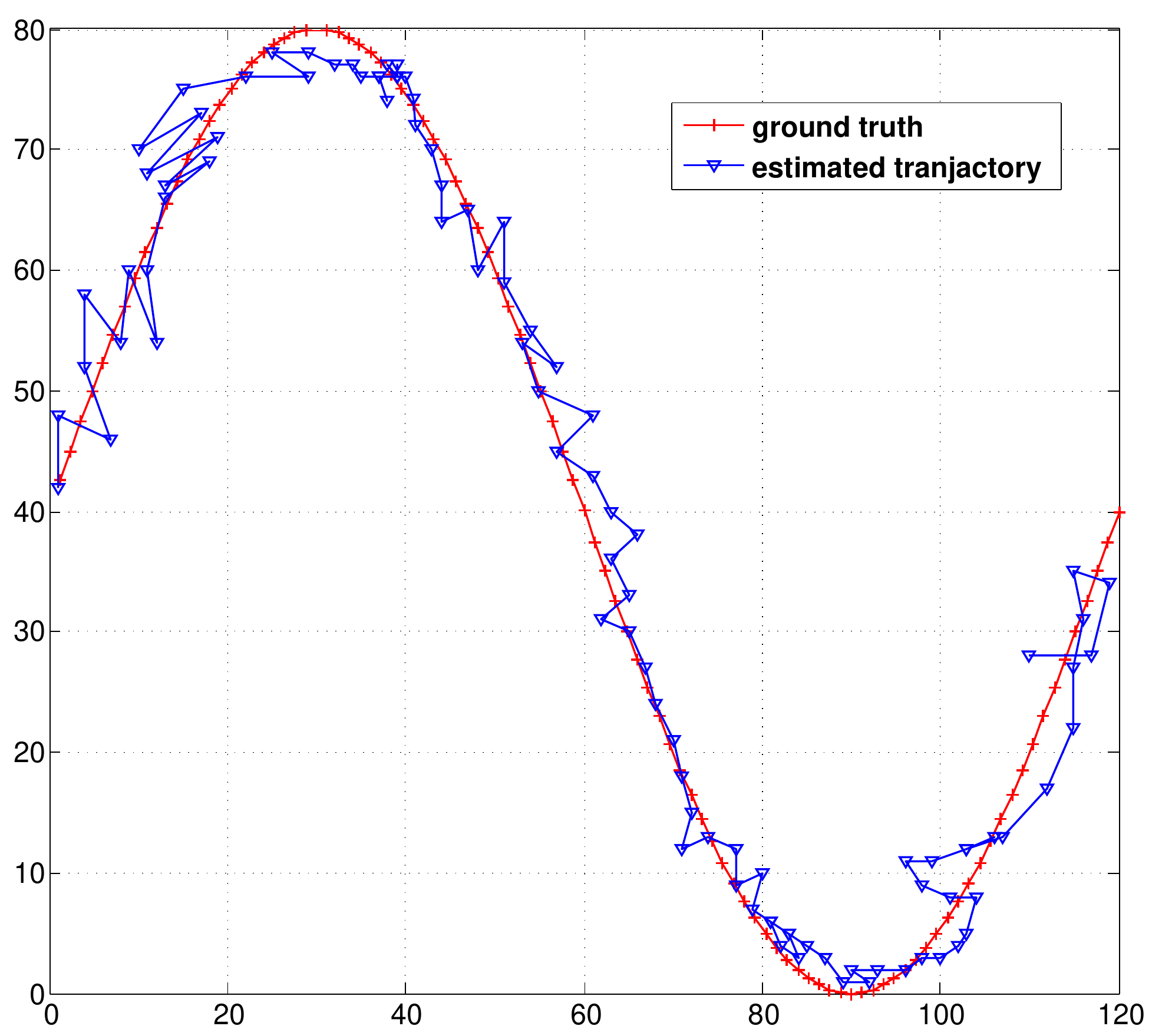} \\
    (a) Hybrid Radio-map with Particle Filter
  \end{minipage}%
    \begin{minipage}[t]{0.33\linewidth} 
    \centering 
    \includegraphics[width=1.8in]{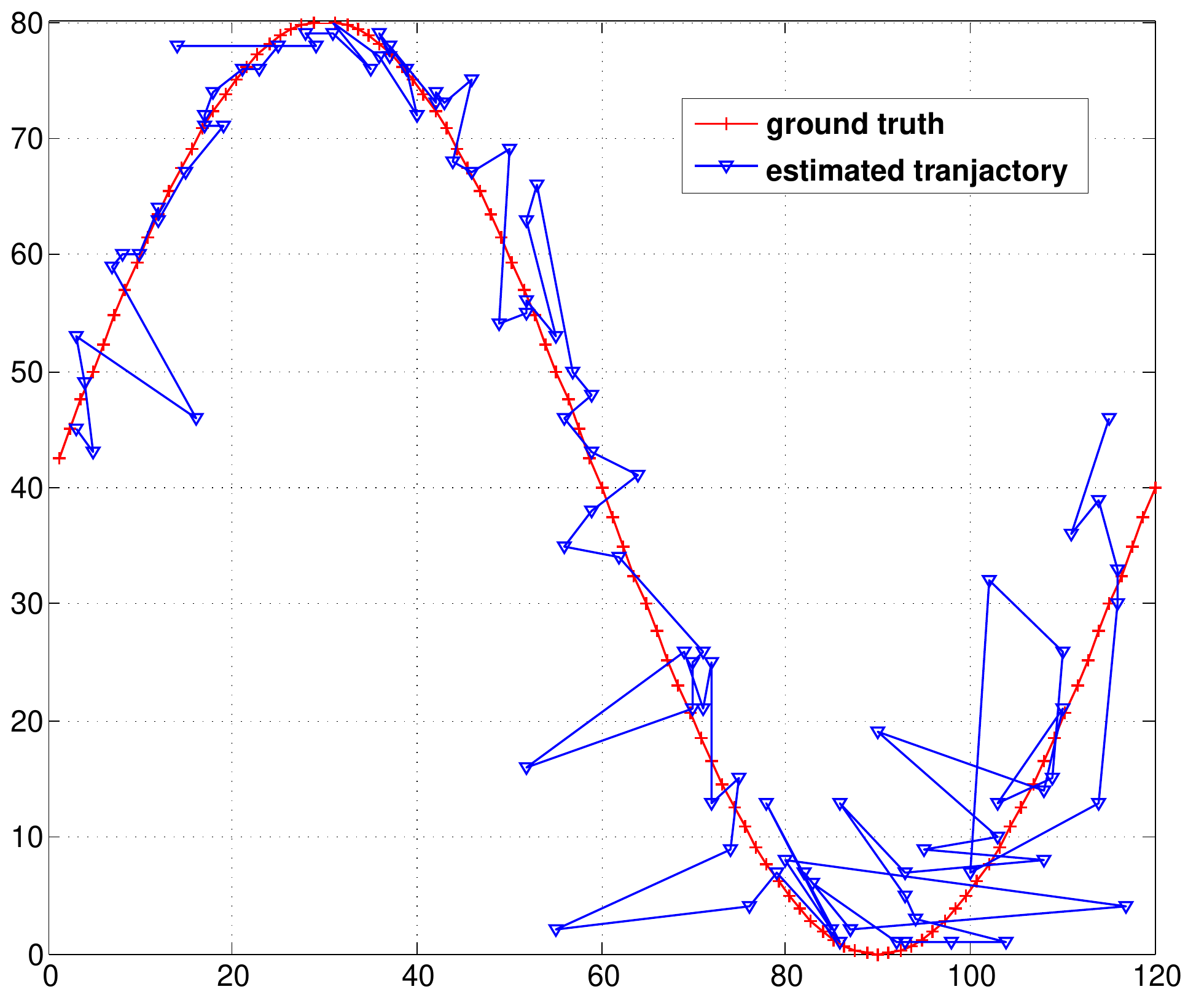} \\
   (b) Hybrid Radio-map without Particle Filter
  \end{minipage}%
    \begin{minipage}[t]{0.33\linewidth} 
    \centering 
    \includegraphics[width=1.8in]{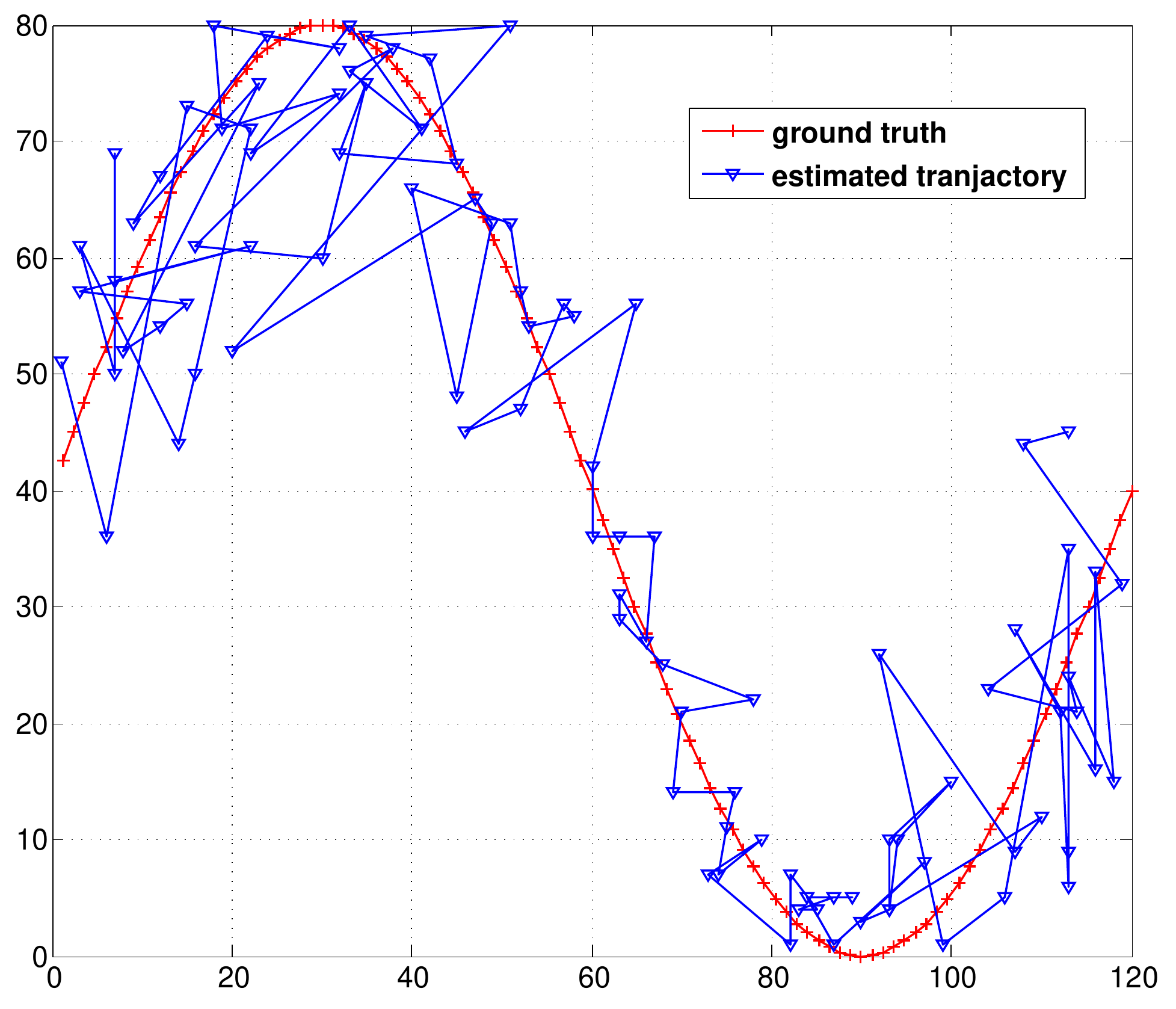} \\
    (c ) RSS-based Radio-map without using TOA signature and Particle Filter
  \end{minipage}%
    \caption{Comparing tracking performances of different algorithms}
    \label{fig6}
\end{figure*}
\section{Simulation and Numerical Results} 
We conducted extensive simulations to verify the advantages of using hybrid radio-map and particle filter than than the traditional RSS-based radio-map method. 
\subsection{Simulation Settings}
The simulation is conducted in Matlab 2012. We provide the code online at \cite{yc}. We simulate the an environment of 120m*80m, in which 10 RSS beacons are randomly deployed. The radio propagation model used in simulation is \cite{lo_adaptive_2012}:
\begin{equation}
{P_r}(d) = {P_t} - {P_l}({d_0}) - 10\eta {\log _{10}}\left( {\frac{d}{{{d_0}}}} \right) + N\left( {0,\sigma } \right)
\end{equation}
We choose $P_{t}$ = 100dbm, $d_{0}=1m$, $\eta=3$, and   $\sigma=3$ in the following simulation results.  Note that since $\sigma=3$,  the variance of RSS can be more serious than general indoor RSS models \cite{bahl_radar:_2000}. 
 
Six TOA beacons are deployed in grid topology. They are activated only when TOA signature is used. The communication radius of each TOA beacon is set to 25m, so that each point in the sensing field can almost be covered by one TOA beacon.  In the offline phase, the training points are selected in $1m*1m$ granularity. Hybrid or RSS radio-maps are trained respectively based on the positioning algorithms.  In online phase, a target moves in the sensing field following a sin-wave path. 
\begin{equation}
\left\{ {\begin{array}{*{20}{l}}
{x = \frac{W}{T}t}\\
{y = \frac{H}{2}\left( {\sin \left( {\frac{{2\pi x}}{W}} \right) + 1} \right)}
\end{array}} \right.
\end{equation}
$W $ and $H$ are the width and height of the sensing field (equal to 120 and 80 respectively in our setting). $T$ is the length of the simulation , so that the target can finish a sin-wave in period $T$. 
 
We evaluated and compared three kinds of positioning algorithms:
\begin{enumerate}
\item \emph{Hybrid radio-map with particle filter}, in which TOA beacons are activated. Fine-grained hybrid radio-map are trained offline (in 1m*1m granularity) and location algorithms introduced in Section III-B, Section III-C are evaluated. 
\item \emph{Hybrid radio-map without particle filter.}  TOA beacons are activated. RSS radio-map are trained offline in 1m*1m granularity. Location algorithm introduced in section III-B is used by limiting the number of positioning candidates to 1, without using particle filter.  
\item \emph{RSS radio-map without particle filter. } TOA beacons are inactivated. RSS radio-map are trained offline in 1m*1m granularity. The position with the least RSS signature difference to the online measured RSS is estimated as position of the target. 
\end{enumerate}
\subsection{Effectiveness of Hybrid Radio-map and Particle Filter}
We at first visually  show the effectiveness of using hybrid radio signature. 
\subsubsection{Narrow down feasible region} 
\begin{figure}[htbp]
\centering
  \begin{minipage}[t]{0.5\linewidth} 
    \centering 
    \includegraphics[width=1.5in]{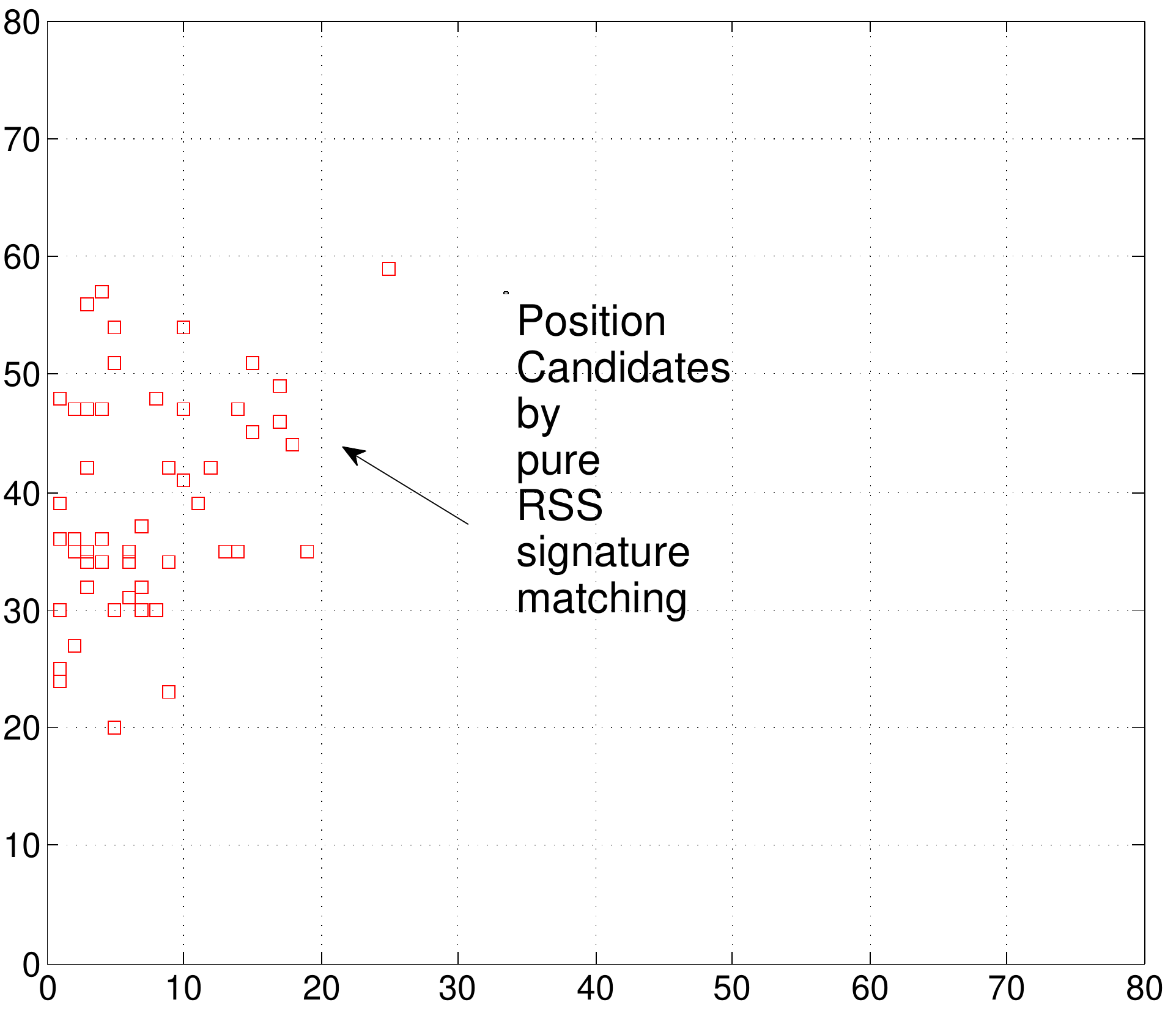} 
  \end{minipage}%
    \begin{minipage}[t]{0.5\linewidth} 
    \centering 
    \includegraphics[width=1.5in]{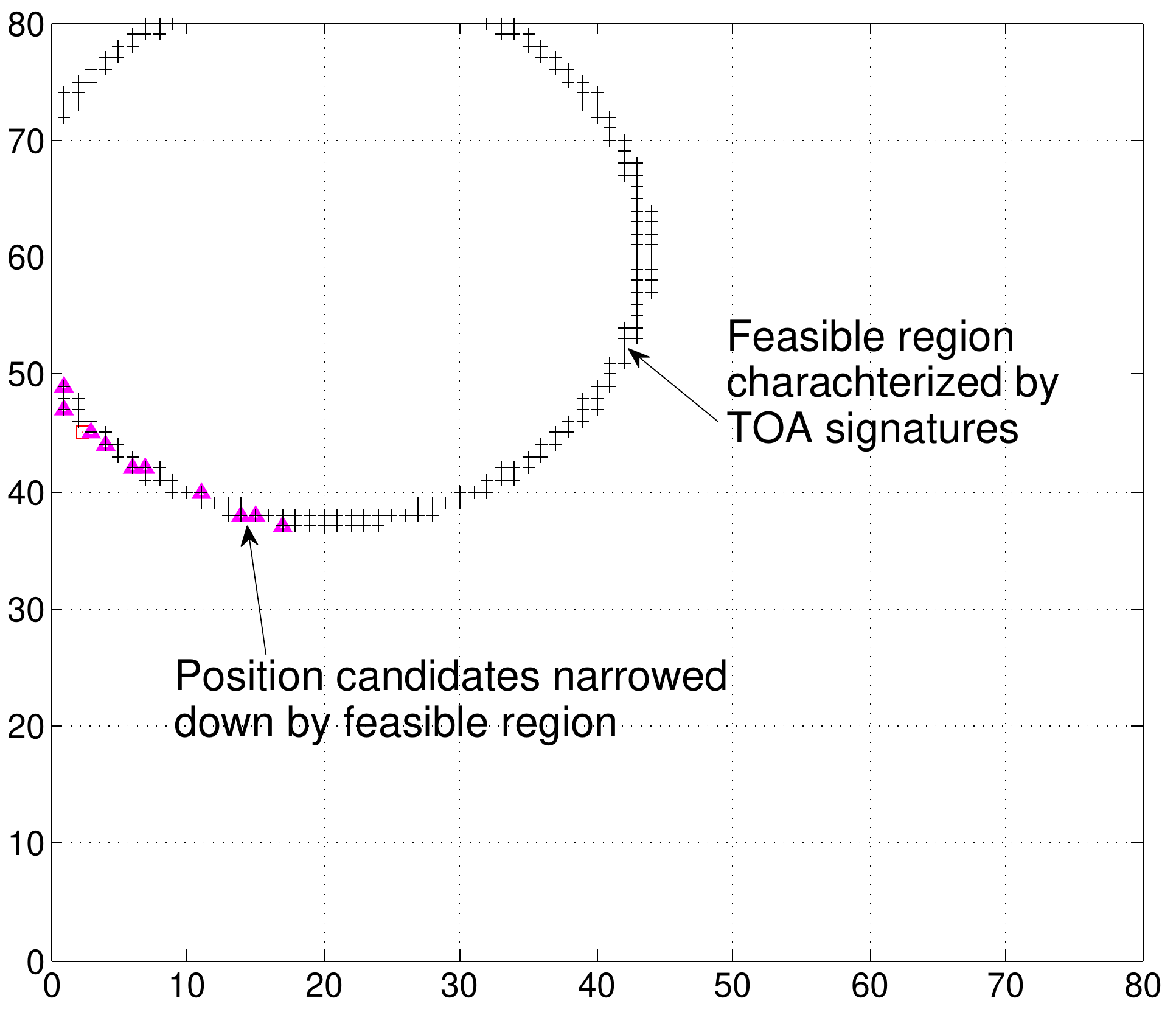} 
  \end{minipage}%
\caption{Search space narrow down by TOA signature}
    \label{fig7}
\end{figure}

One important contribution of the TOA signature is to provide low variance position estimation and dramatical search space narrowing down.  As shown in Fig.\ref{fig7}a, the red square points are possible position candidates  generated by least RSS-distance. We can see the estimated possible positions are highly scattered in the sensing field for the unreliability of the RSS signal. Fig.\ref{fig7}b shows how the TOA-signature helps to narrow down the feasible region. Since the possible positions must be in the feasible region (the circle area), the possible positions are filtered.   

\subsubsection{Improve the positioning accuracy}
Fig.\ref{fig6} compares different algorithms to show the improvement of positioning accuracy by the hybrid radio-map and particle filter algorithms. The red sin-wave  curves in the figures shows the ground truth of the target movement.  Fig.\ref{fig6}(a) illustrates the target tracking performance when using hybrid radio-map plus particle filter tracking algorithm. The algorithm can provide accuracy target tracking performance. Fig.\ref{fig6}(b) shows the tracking results when only hybrid radio-map is used without the particle filter algorithm.  Instead of particle filter, at each step, the candidate position which matches best to the hybrid signature of measurement is chosen as the position estimation. We can see the positioning performance degrades much than that in Fig.\ref{fig6}(a). Fig.\ref{fig6}c shows the tracking results when only RSS-based radio-map is used without using TOA signature nor particle filter. At each step, the candidate position in the radio-map which matches best to RSS-signature of target is chosen as the position estimation. We can see that the tracking performances become much worse than the prior two approaches. 
\begin{figure}[htbp]
\begin{center}
\includegraphics[width=2.2in]{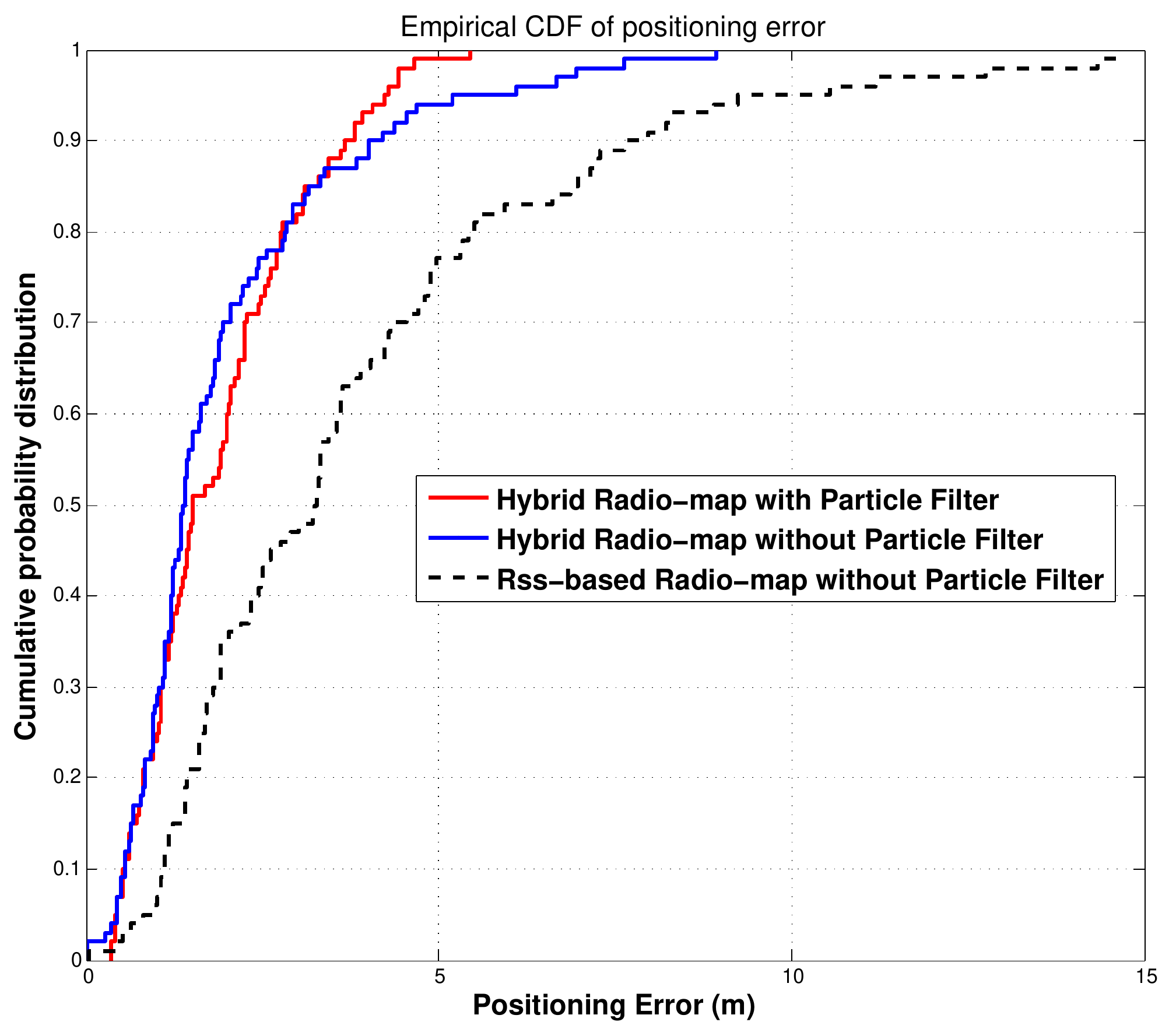}
\caption{Cumulative probability distribution of positioning errors for different positioning algorithms.}
\label{cdf}
\end{center}
\end{figure}

Fig.\ref{cdf} uses cumulative probability distribution of positioning error to further illustrate the locating performance improvement achieved by hybrid radio-map and particle filter  algorithms.  The results are based on the average positioning errors of 20 simulations. It show that the method of hybrid radio-map plus particle filter performs the best, which is a little better than the method of using hybrid radio-map but without particle filter. The difference of them is that the latter method may has a small portion of results having large positioning error, which is not robust. Both of these two methods are much better than the traditional methods using RSS-based radio-map.  The results show the significance of the sparse low-variance signature for the improvement of positioning accuracy. 

\section{Conclusion}
This paper presents hybrid radio-map method for improve the positioning accuracy of RSS-based wireless indoor localization. It presents efficient methods to utilize the sparse, low variance signature to construct hybrid radio-map, and presents particle filter  based algorithm for accurate online target tracking. Simulation results verified that by using very limited TOA beacons, the hybrid radio-map method can dramatically improve the positioning accuracy of wireless location systems. We will conduct hardware experiments and system development in our future work. 

\section*{Acknowledgment}
This work was supported in part by National Natural
Science Foundation of China Grant 61202360, Major Science and Technology Project on Transportation Informatics for  Communication Networks of Internet of Ships (2012-364-222-203), and the National Basic Research Program of China Grant 2011CBA00300,
2011C-BA00302.



\bibliographystyle{IEEEtran}
\bibliography{IEEEabrv,./ref.bib}
%


\end{document}